\documentclass{ws-ijbc}
\usepackage{ws-rotating}     

\begin{document}

\catchline{}{}{}{}{} 

\markboth{Author's Name}{Paper Title}

\title{Orbits in a non-Kerr Dynamical System}

\author{G.~Contopoulos}

\address{Research Center for Astronomy, Academy of Athens \\
Soranou Efesiou 4, GR-11527, Athens, GREECE \\
gcontop@academyofathens.gr}

\author{G.~Lukes-Gerakopoulos}

\address{Research Center for Astronomy, Academy of Athens \\
Soranou Efesiou 4, GR-11527, Athens, GREECE \\
gglukes@phys.uoa.gr}

\author{T.~A.~Apostolatos}

\address{Section of Astrophysics, Astronomy, and Mechanics, \\
Department of Physics, University of Athens\\ Panepistimiopolis
Zografos GR15783, Athens, Greece\\
thapostol@phys.uoa.gr}

\maketitle

\begin{history}
\received{(to be inserted by publisher)}
\end{history}

\begin{abstract}

We study the orbits in a Manko-Novikov type metric (MN) which is a perturbed
Kerr metric. There are periodic, quasi-periodic, and chaotic orbits, which are
found in configuration space and on a surface of section for various values of
the energy $E$ and the $z-$component of the angular momentum $L_z$. For
relatively large $L_z$ there are two permissible regions of non-plunging motion
bounded by two closed curves of zero velocity (CZV), while in the Kerr metric
there is only one closed CZV of non-plunging motion. The inner permissible
region of the MN metric contains mainly chaotic orbits, but it contains also a
large island of stability. When $L_z$ decreases the two permissible regions join
and chaos increases. Below a certain value of $L_z$ most orbits escape inwards
and plunge through the horizon. On the other hand as the energy $E$ decreases
(for fixed $L_z$) the outer permissible region shrinks and disappears. In the
inner permissible region chaos increases and for sufficiently small $E$ most
orbits are plunging. We find the positions of the main periodic orbits as
functions of $L_z$ and $E$, and their bifurcations. Around the main periodic
orbit of the outer region there are islands of stability that do not appear in
the Kerr metric (integrable case). In a realistic binary system, because of the
gravitational radiation, the energy $E$ and the angular momentum $L_z$ of an
inspiraling compact object decrease and therefore the orbit of the object is
non-geodesic. In fact in an extreme mass ratio inspiraling (EMRI) system the
energy $E$ and the angular momentum $L_z$ decrease adiabatically and therefore
the motion of the inspiraling object is characterized by the fundamental 
frequencies which are drifting slowly in time. In the Kerr metric the ratio of
the fundamental frequencies changes strictly monotonically in time. However, in
the MN metric when an orbit is trapped inside an island the ratio of the 
fundamental frequencies remains constant for some time. Hence, if such a
phenomenon is observed this will indicate that the system is non integrable and 
therefore the central object is not a Kerr black hole.
  
\end{abstract}


\section{Introduction}

The geodesics in a Kerr metric are derived from an integrable system of
equations. The Kerr metric itself is characterized by its mass $M$ and spin $S$.
However, there are other solutions of the vacuum Einstein equations, that are
close to the Kerr solution, in which the geodesic equations of motion are
nonintegrable. The geodesic motions in such backgrounds are either ordered or
chaotic. It would be of great interest to check whether the massive compact
objects that lie at the center of galaxies are Kerr black holes or some other
type of exotic objects (see e.g. \cite{Johan10a,Johan10b}). This could be
attained by analyzing the gravitational waves  emitted by a compact object (of
mass $1-10^2~M_\odot$) inspiraling around the central massive object (with mass
$10^5-10^9~M_\odot$) that lies at the center of a galaxy; such binary systems
are called EMRIs (Extreme Mass Ratio Inspirals). It is expected that future
low-frequency detectors of gravitational waves, like the space detector LISA
\cite{LISA}, will provide us sufficient amount of information, by analyzing the
spectrum of the gravitational-wave signal, to answer the question whether the
central body is a Kerr black hole or any other kind of a non-Kerr object.

In order to make this distinction more clear we should analyze in detail all
types of geodesic orbits in a generic non-Kerr background and focus mainly on
the qualitative differences from the corresponding orbits in a Kerr system.

Although there is a variety of vacuum solutions of Einstein equations that could
be used as background metrics to study non-Kerr geodesics we have used a
specific family of asymptotically flat spacetimes that incorporates the basic
characteristics of a generic metric that could be formed as a deviation from the
Kerr metric. The particular family MN is a one-parameter subfamily of the
so-called Manko-Novikov multiparametric family of metrics \cite{Manko92}. The
Manko-Novikov family of spacetimes was actually constructed to describe the
exterior vacuum of any axisymmetric object one could consider. By setting the
value of the new parameter of MN equal to zero we get back the Kerr metric and
that is why the MN metric was named by Gair etal \cite{Gair08} a ``bumpy black
hole spacetime (it should be emphasized though that there are infinite ways to
create bumpy black hole spacetimes, and MN is simply one of those). The orbits
in this metric were first studied in \cite{Gair08} and more thoroughly later in
\cite{Apostolatos09,Lukes10}. In the present paper we study in a more systematic
way the orbits in a MN metric for a great variety of parameter values and we
discuss their effects on the spectrum of the corresponding gravitational waves.

The form of the MN metric is given in \cite{Gair08,Lukes10}. We do not reproduce
it here because it is given by a rather long and complicated formula, which is
used by our numerical code in order to solve the geodesic equations for a test
particle in the corresponding background.

The MN metric is assumed to have the same mass $M$ and spin $S$ as the
corresponding Kerr metric, while its quadrupole moment 
\begin{equation}
 M_2=-M\left[\left(\frac{S}{M}\right)^2+q~M^2\right]
\end{equation}
differs from the corresponding Kerr moment $M_{2,K}=-S^2/M$ by the quantity
$-q~M^3$, where $q$ is the new parameter of the MN metric that produces the
deviation of the MN metric from the corresponding Kerr metric. Actually all
higher than the quadrupole multipole moments are different from the
corresponding multipole moments of Kerr as well when $q$ is non-zero. If $q=0$
the MN metric is reduced to Kerr. In the present paper we consider only the case
$q>0$, which corresponds to an oblate perturbation of the Kerr metric.

In section \ref{sec:OrbMN} we study systematically the orbits for a wide range
of parameter values of the MN metric. In section \ref{sec:Charact} we study the 
characteristics of the main periodic orbits. In section \ref{sec:NGO} we focus
our attention on the effects of the non geodesic orbits on the corresponding
gravitational waves through which one could in principle distinguish a non-Kerr
from a Kerr metric. Finally in section \ref{sec:Conc} we draw our conclusions.

\section{Orbits in the Manko-Novikov (MN) metric} 
\label{sec:OrbMN}

\begin{figure}[ht]
\begin{center}
\psfig{file=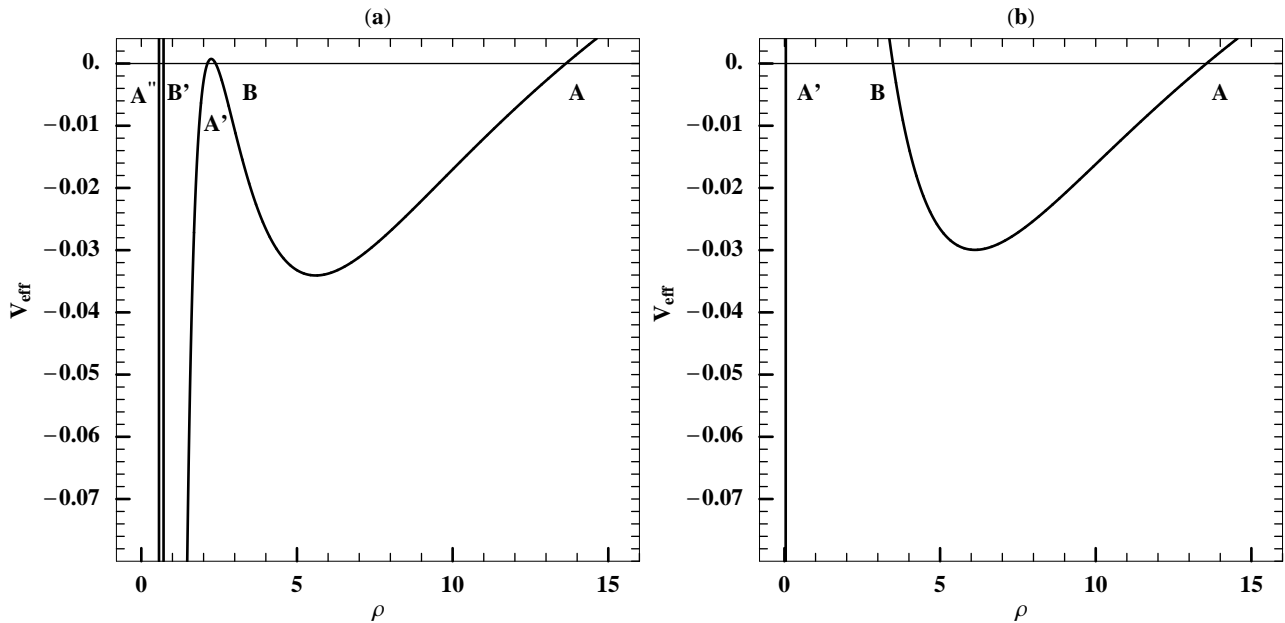,width=7in} 
\end{center}
 \caption{(a) The effective potential $V_{\rm{eff}}$ as a function of $\rho$,
 for $z=0$ (on the equatorial plane) and
 $E=0.95$, $L_z=3$, $q=0.95$, $\chi=0.9$, $M=1$. (b) The corresponding 
 $V_{\rm{eff}}$ in the Kerr case ($q=0$).}
\label{FigVeff}
\end{figure}

In the framework of General Relativity the Einstein field equations play the
role of the classical Poisson equation for a given mass and energy distribution.
The metric of the spacetime induced by such a distribution determines the
geodesics followed by the test particles, in analogy with the trajectories
followed by test particles in a given ``classical'' gravitational potential. One
of the well-known and astrophysically relevant vacuum solutions of the Einstein
field equations is the Kerr metric, which is characterized uniquely by a mass
and an angular momentum. A broader family of vacuum solutions of Einstein
equations, known as Manko-Novikov (MN) solutions, depend on one more parameter
than Kerr, the quadrupole-deviation parameter $q$. By setting $q=0$ we obtain
the Kerr solution. This new metric can be used to describe approximately the
exterior of an axisymmetric object that has a finite distribution of mass and
angular momentum, and it would be crucial if we could somehow detect qualitative
differences in the geodesics between the two types of metrics.

In the following we use the Weyl-Pappapetrou line element expressed in
cylindrical coordinates $(t,\rho,~z,~\phi)$ to describe the orbits (see
e.g.~\cite{Gair08} or \cite{Lukes10}). The MN system has an axis of symmetry $z$
and a plane of symmetry $z=0$. Thus we consider orbits on the meridian plane
$(\rho,~z)$, while the azimuth $\phi$ can be easily computed from $\rho(\tau)$
and $z(\tau)$  (see \cite{Lukes10}), where $\tau$ is the proper time.

The MN is a stationary axisymmetric metric, thus the geodesic orbits are
characterized by two integrals of motion: the energy $E=-p_t$ and the
z-component of the angular momentum $L_z=p_z$. The test mass $\mu$ of the 
orbiting body is also fixed and is represented by the constancy of the 
Hamiltonian function itself. In order to study various types of geodesic orbits
we have assumed fixed values for the mass ($M=1$) and the spin ($S=0.9~M^2$) of 
the metric,while we have chosen various values for the orbital parameters
$E,L_z$ and the quadrupolar parameter $q$. By choosing $M=1$ the dimensionless
spin parameter $\chi=S/M^2$ is equal to the spin $S$ itself.

The velocities $\dot{\rho}$ and $\dot{z}$ satisfy an equation analogous to a 
generic 2-dimensional Newtonian problem under conservative forces
\begin{equation} \label{EqClVeff}
\frac{1}{2}(\dot{\rho}^2+\dot{z}^2)+V_{\rm{eff}}(\rho,z)=0
\end{equation}
where $V_{\rm{eff}}$ is an effective potential that depends on the coordinates
$\rho,~z$, the constants of motion $E,~L_z$, and the parameters of the metric 
$M,~S,~q$ (see \cite{Lukes10} for details). A test particle is allowed to move
only in the regions where $V_{\rm{eff}}\leq 0$. The boundaries of these regions 
are the so called CZVs (curves of zero velocity) since there the orbits assume 
zero velocity ($\dot{z}=\dot{\rho}=0$). For $z=0$ (along the equatorial plane) 
the effective potential $V_{\rm{eff}}$ as a function of $\rho$ has the
form shown in Fig. \ref{FigVeff}a. The corresponding allowed region for an orbit 
is either between $A$ and $B$, or between $A^\prime$ and $B^\prime$, or between 
$A^{\prime\prime}$ and $\rho=0$. The corresponding effective potential in the 
Kerr case is shown in Fig. \ref{FigVeff}b. In the Kerr case $\rho$ can vary 
either between $A$ and $B$, or between $A^{\prime}$ and $\rho=0$. The allowed 
regions that include the $\rho=0$ point correspond to plunging orbits, that is
to orbits that will end up plunging to the central singularity of the metric.

\begin{figure}[ht]
\begin{center}
\psfig{file=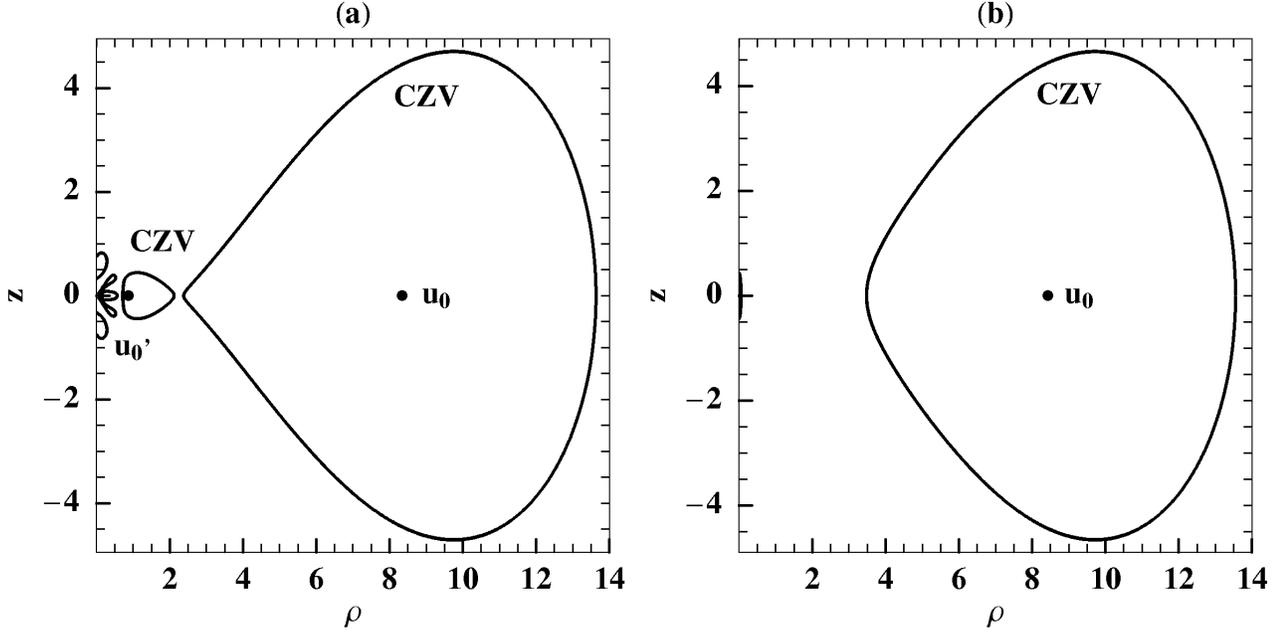,width=7in} 
\end{center}
 \caption{(a) The permissible regions (inside the CZVs) of the motion on the
 meridian plane $(\rho,z)$ for the parameters $E=0.95$, $L_z=3$, $q=0.95$,
 $\chi=0.9$ and $M=1$. (b) The permissible region in the corresponding Kerr case
 ($q=0$; all other parameters as in (a)).}
\label{FigConLz3E095}
\end{figure}

\begin{figure}[ht]
\begin{center}
\psfig{file=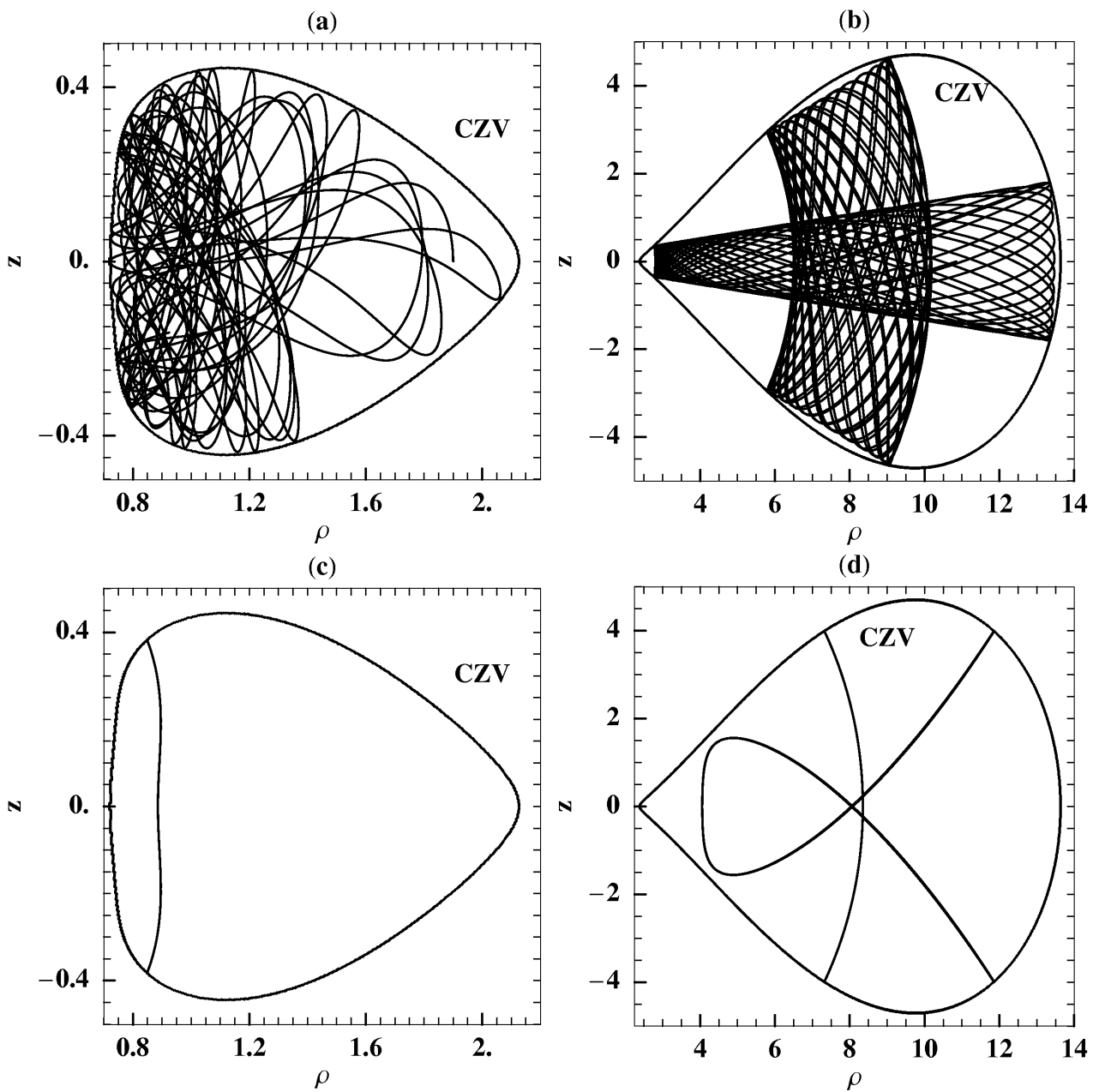,width=7in} 
\end{center}
 \caption{(a) A chaotic orbit inside the inner CZV of Fig.~\ref{FigConLz3E095}a.
  (b) Two ordered orbits inside the outer CZV of Fig.~\ref{FigConLz3E095}a.
  (c) The periodic orbit starting from the point $\mathbf{u}^\prime_0$. 
  (d) The periodic orbit starting from the point $\mathbf{u}_0$, and a periodic
  orbit of multiplicity 3 on the surface of section $z=0$. Note that only the
  edge points of the latter orbit touch the CZV.}
\label{FigConOrgCha}
\end{figure}

For a certain range of parameters there are two distinct  closed CZVs within
which there are bound geodesic orbits. The central points $\mathbf{u}_0$
($\rho\approx 8.35,~z=0$) and $\mathbf{u}^\prime_0$ ($\rho \approx 0.886,~z=0$),
corresponding to the values of the parameters shown in Fig.~\ref{FigConLz3E095}a,
represent the initial conditions (along $\dot{\rho}=0$) for the two main
periodic orbits shown in Figs.~\ref{FigConOrgCha}c,d. These orbits oscillate
periodically around the equatorial plane $z=0$. Close to $\rho=0$ there are five
more CZV curves that  are in contact with the neighborhood of $\rho=0$. Orbits
inside these CZVs eventually plunge through the horizon which lies along the
segment $\rho=0$, $|z| \leq k$,  where 
\begin{equation}
k = M
\frac{\chi^2-(\sqrt{\chi^2-1}-\chi)^2}{\chi^2+(\sqrt{\chi^2-1}-\chi)^2} 
\end{equation}
The horizon of the MN metric is not fully enclosing the central singularity.
Therefore the fact that the MN metric is not characterized only by its mass and
spin, but also by its oblateness $q$, is compatible with the no-hair theorem.
In fact the horizon of MN is cut by a line singularity across the equatorial
line $\rho=z=0$ \cite{Gair08}, while this singularity does not exist in the case
of the Kerr metric. As mentioned before the Kerr metric has only one closed
CZV around the point $\mathbf{u}_0$ $(\rho\approx 8.44,~z=0$ corresponding to
the values of the parameters shown in Figs.~\ref{FigConLz3E095}a,b). The
$\mathbf{u}_0$ point marks the initial condition (with $\dot{\rho}=0$) for the
periodic orbit in Kerr.

\begin{figure}[ht]
\begin{center}
\psfig{file=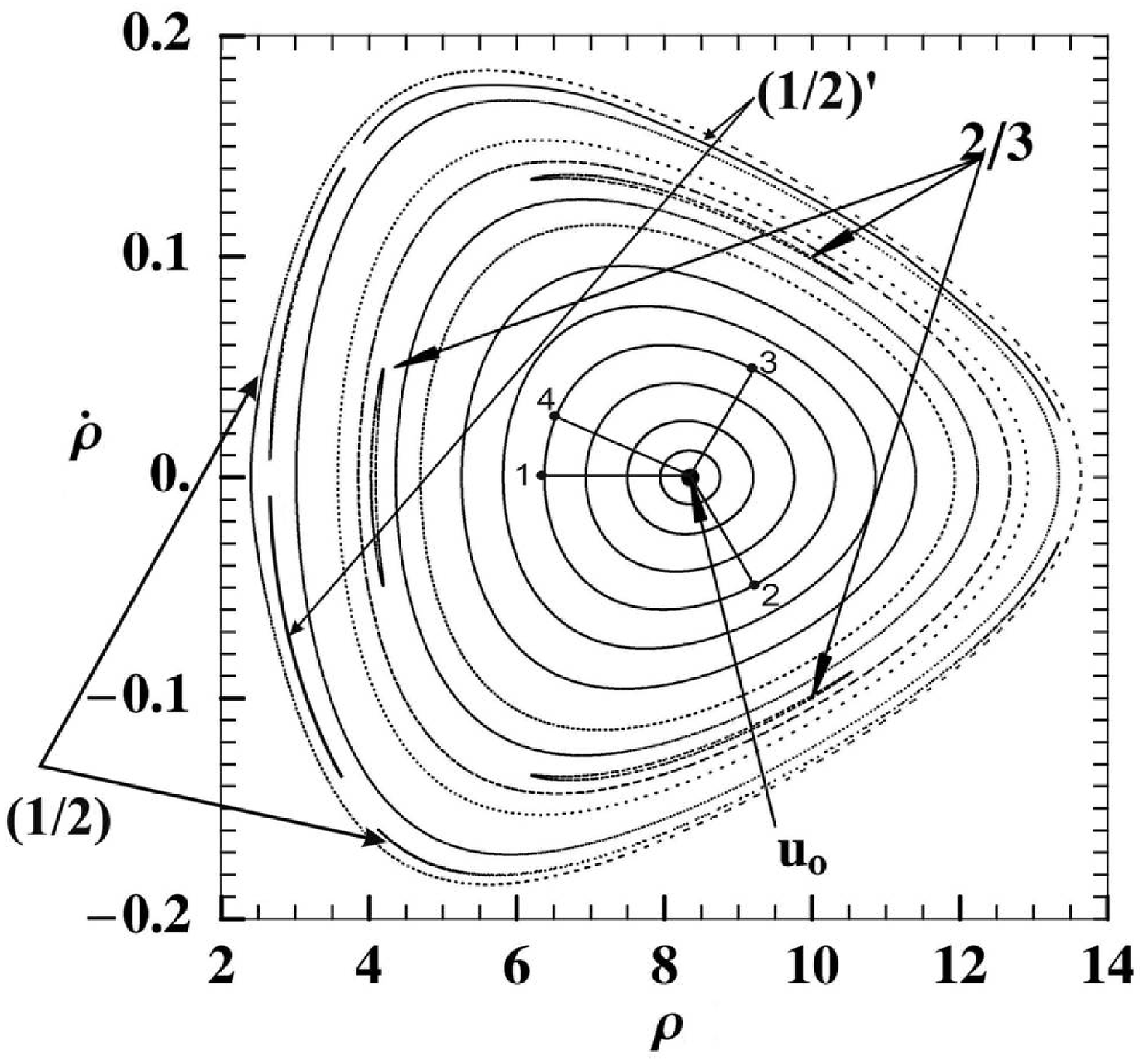,width=3.5in} 
\end{center}
 \caption{The Poincar\'{e} surface of section ($z=0$) in the MN case containing
 orbits of the outer CZV.}
\label{FigPSLz3E095}
\end{figure}

Inside the outer CZV of the MN metric most orbits are ordered
(Fig.~\ref{FigConOrgCha}b), while in the inner CZV most orbits are chaotic
(Fig.~\ref{FigConOrgCha}a). A method to distinguish between ordered and chaotic
orbits is by considering a Poincar\'{e} surface of section, i.e. a surface
intersecting all the orbits in phase space and finding the successive
intersections of every orbit by this surface. The successive intersections of an
ordered orbit are along a curve that is called invariant curve
(Fig.~\ref{FigPSLz3E095}), while the successive iterates of a chaotic orbit are
scattered irregularly. The invariant curves either encircle the central periodic
orbit (the one that passes through the point $\mathbf{u}_0$ with $\dot{\rho}=0$)
or form a chain of islands of stability around stable resonant periodic orbits.
In Fig. \ref{FigPSLz3E095} we see 3 islands of stability around a resonant orbit
(resonance $2/3$), and 2 couples of islands around 2 different resonant periodic
orbits of the resonance $1/2$. Between the 3 islands of stability there are 3
points corresponding to an unstable periodic orbit of the $2/3$ resonance.
Between the 4 islands of stability there are 4 points corresponding to 2
unstable periodic orbits of the resonance $1/2$, located at symmetrical
positions with respect to the axis $z=0$. Near the unstable periodic orbits the
orbits are chaotic and their iterates on the Poincar\'{e} surface of section are
scattered. In the case of Fig.~\ref{FigPSLz3E095} this scatter lies in extremely
thin zones around and between the islands of stability. These zones are not
visible in the figure.

\begin{figure}[ht]
\begin{center}
\psfig{file=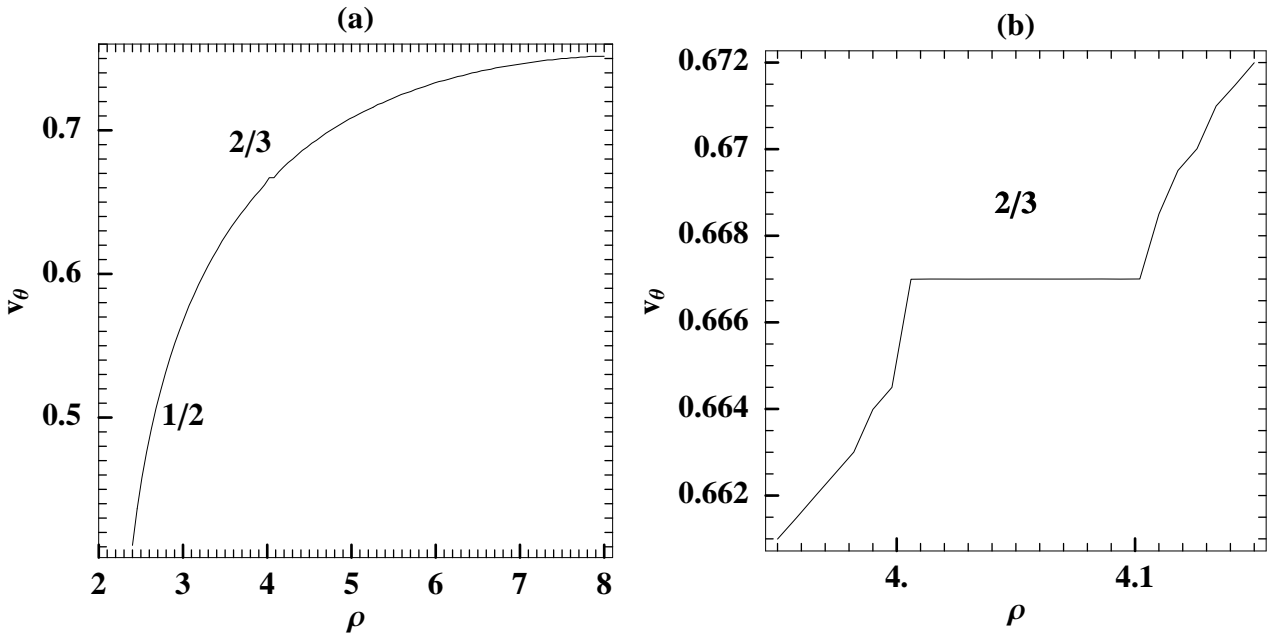,width=7in} 
\end{center}
 \caption{(a) The rotation number $\nu_\theta$ as a function of $\rho$  along
 the line $\dot{\rho}=0$ (and $z=0$, $\dot{z}>0$). (b) A magnification of the
 region close to the plateau at $\nu_\theta=2/3$.}
\label{FigRotLz3E095}
\end{figure}

In order to find the islands of stability in Fig.~\ref{FigPSLz3E095} it is
necessary to proceed along small steps in moving from one invariant curve to the
next one. A systematic way to find the resonant islands and the respective
chaotic domains is by calculating the rotation numbers along successive orbits
at various distances from the central orbit $\mathbf{u}_0$ (for a review see
\cite{Contop02}). Thus in Fig.~\ref{FigPSLz3E095} we have drawn the first 4
iterates of a regular orbit that has a rotation number larger than $2/3$. In
order to measure the rotation number we compute the average fraction of a circle
as we advance from one iterate to the next moving clockwise. The rotation number
$\nu_\theta$, as a function of $\rho$ along the axis $\dot{\rho}=0$ (on the
surface of section at $z=0$) looks like a smooth monotonic curve
(Fig.~\ref{FigRotLz3E095}a). This is actually a smooth strictly monotonic 
curve in an integrable case, like the Kerr metric. However, in a nonintegrable
case there are many (in fact infinite) plateaus near every resonant value 
(rational number), like the plateau of Fig.~\ref{FigRotLz3E095}b. In order to
find these plateaus we require a small step $\Delta \rho$ between successive
initial values of $\rho$. The orbits corresponding to the initial conditions 
of a plateau form a chain of islands around the stable resonant orbit, all with
the same rotation number as the periodic orbit that lies at the centers of these
islands, e.g.~the orbits in the plateau $2/3$ form $3$ islands, all having the
same rotation number $2/3$.

In an integrable case, like the Kerr metric, there are no islands of stability.
All the resonant periodic orbits are neutrally stable and do not form islands
around them. In fact a resonant invariant curve on which lies a periodic orbit
of period $T$  has all its points as initial conditions of periodic orbits with
the same period $T$. Thus the existence of islands of stability is an indication
of non-integrability. However, there are particular integrable systems with one
type of islands of stability \cite{Contop78}, e.g.~a Hamiltonian system of two
degrees of freedom expressed in action angle variables which depends on the two
actions $J_1,~J_2$ and a particular combination of the angles
$n~\theta_1-m~\theta_2, \textrm{with}~n,~m \in \mathbb{N}$. Nonetheless a system
with more than one type of islands (e.g.~both double and triple islands in the
same system) cannot be integrable. This is consistent with the non-existence of
a Carter-type integral in generic non-Kerr cases \cite{Flanagan07,Brink08}.
However the outer permissible region of the MN system (the surface of section of
which has been drawn in Fig.~\ref{FigPSLz3E095}), is quite close to an
integrable one. In this system chaos is very limited and thus not visible in
Fig.~\ref{FigPSLz3E095}. It exists mainly close to the unstable periodic orbits
that lie between the islands of each chain of islands.

\begin{figure}[ht]
\begin{center}
\psfig{file=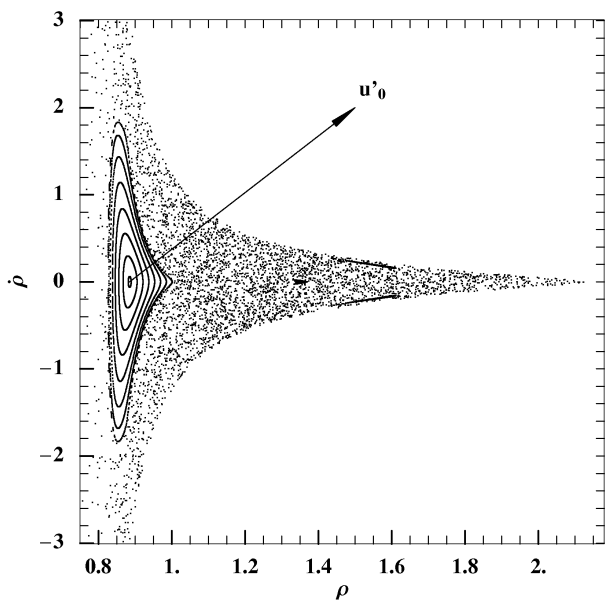,width=3.2in} 
\end{center}
 \caption{The Poincar\'{e} surface of section of the inner permissible region
in the case of Fig.~\ref{FigConLz3E095}a}.
\label{FigIntReg}
\end{figure}

On the other hand, the inner CZV (Fig.~\ref{FigConLz3E095}a) contains many
chaotic orbits, but it contains also some ordered orbits. On the Poincar\'{e}
surface of section (Fig.~\ref{FigIntReg}) we see the chaotic domain and a large
island of stability around a point $\mathbf{u}_0^\prime$. There are also some
very small islands of stability of multiplicity 3. The orbits in the inner
regions of the MN system have not been studied in detail up to now and they have
been considered ``a very interesting puzzle'' \cite{Brink08}. For this reason we
study them in some detail in the present paper.

\begin{figure}[ht]
\begin{center}
\psfig{file=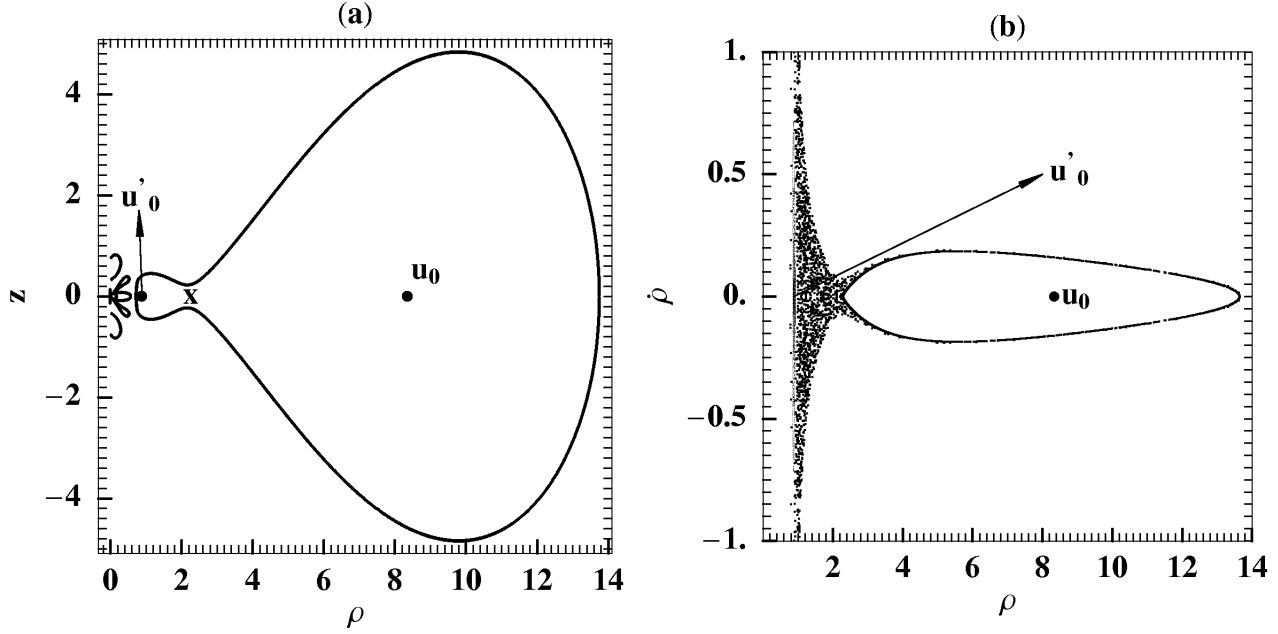,width=7.in} 
\end{center}
 \caption{(a) The CZVs on the $(\rho,z)$ plane for $E=0.95$, $L_z=2.98$ and
 $q=0.95$. The thick dots represent initial conditions  (along with
 $\dot{\rho}(0)=0$) for stable periodic orbits while the ``x'' marks
 the initial condition of an unstable periodic orbit.
 (b) On the surface of section $z=0$ the chaotic domain of the inner permissible
 region expands and forms a chaotic layer just inside the boundary of the outer
 region.}
\label{FigLz2_98}
\end{figure}

If we decrease $L_z$, while keeping $E$ and $q$ fixed, the topology of the CZV
changes (Fig. \ref{FigLz2_98}a). Namely for $L_z=2.98$, just a little smaller
from the value $L_z=3$ of Fig.~\ref{FigConLz3E095}a, the two main CZVs have
joined. At the point of junction an unstable periodic orbit is formed between
the two previously separate CZVs. This saddle point exists for $L_z<2.99761912$
and corresponds to the local maximum of the effective potential $V_{{\rm eff}}$
near $\rho=2$ (Fig.~\ref{FigConLz3E095}). On the corresponding Poincar\'{e}
surface of section the permissible regions are joined; that is there are chaotic
orbits that move in both regions. In this case the chaotic sea of the inner
region has now been extended into the outer region as well, forming a chaotic
layer around the whole ordered region which surrounds the point $\mathbf{u}_0$
(Fig.~\ref{FigLz2_98}b).

\begin{figure}[ht]
\begin{center}
\psfig{file=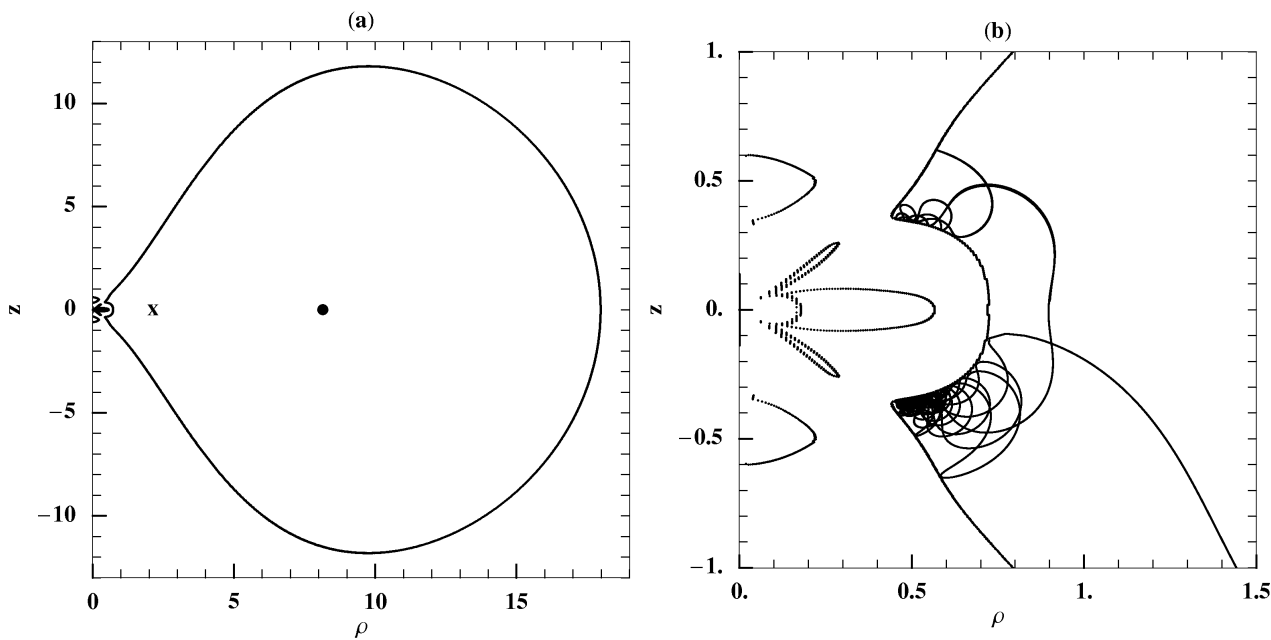,width=7.in} 
\end{center}
 \caption{(a) The CZVs for $L_z=1.7,~E=0.95,~q=0.95$. The dot at the center of
 the diagram represents the initial condition for the stable orbit
 $\mathbf{u}_0$ and the ``x'' represents the initial condition for the unstable
 periodic orbit. (b) A chaotic orbit in the inner part of the permissible space.
  }
\label{FigConArLz1_7}
\end{figure}

As $L_z$ decreases further the neck joining the inner and the outer regions of
the CZV, containing the non-plunging orbits, expands (Fig.~\ref{FigConArLz1_7}a). 
Now the minimum $\rho$ of the former CZV does not lie on the $\rho-$axis ($z=0$),
as in Fig.~\ref{FigLz2_98}a. In fact there are two minima with $z \neq 0$, 
at symmetrical locations with respect to the $\rho-$axis, which define the tips
of two extensions of the CZV each approaching one of the two inner lobe-like
CZVs which contain plunging orbits. These two CZVs lie on either side of the
central lobe-like CZV which also contains plunging orbits. The permissible
region of non-plunging orbits near these extensions contains mostly chaotic
orbits (Fig.~\ref{FigConArLz1_7}b).

\begin{figure}[ht]
\begin{center}
\psfig{file=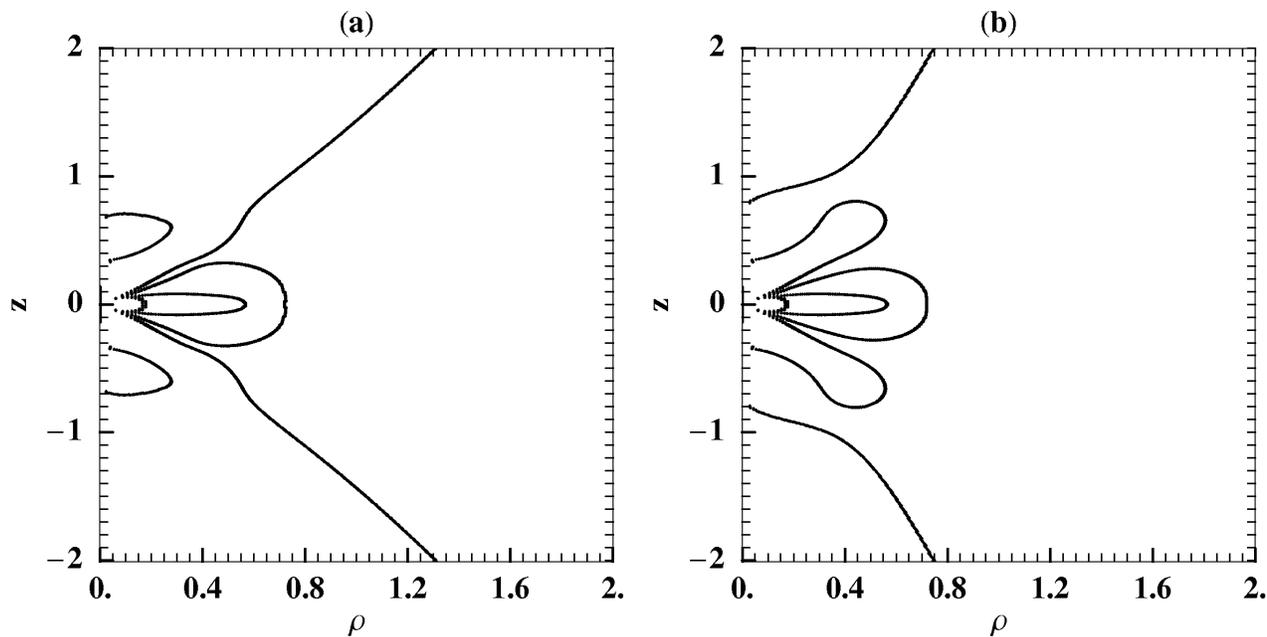,width=7.in} 
\end{center}
 \caption{(a) The CZVs for $L_z=1.6,~E=0.95,~q=0.95$ and (b) the CZVs for
 $L_z=1,~E=0.95,~q=0.95$ near the horizon $\rho=0$.}
\label{FigLzPlung}
\end{figure}

For even lower values of $L_z$ ($L_z=1.6,~E=0.95,~q=0.95$) the aforementioned
extensions join the two symmetrical plunging areas around the central one 
(Fig.~\ref{FigLzPlung}a) and all the chaotic orbits belonging to the
chaotic sea, surrounding the island around the stable orbit $\mathbf{u}_0$, 
eventually plunge through the horizon. In Fig.~\ref{FigLzPlung}a we observe
the formation of another two symmetrical extensions of the main CZV for $|z|$
values a little higher than for the already joined extensions. These extensions
for still smaller $L_z$ values ($L_z=1$) join the two more distant plunging
permissible areas (Fig.~\ref{FigLzPlung}b) and create two more channels for
chaotic orbits to plunge in.

\begin{figure}[ht]
\begin{center}
\psfig{file=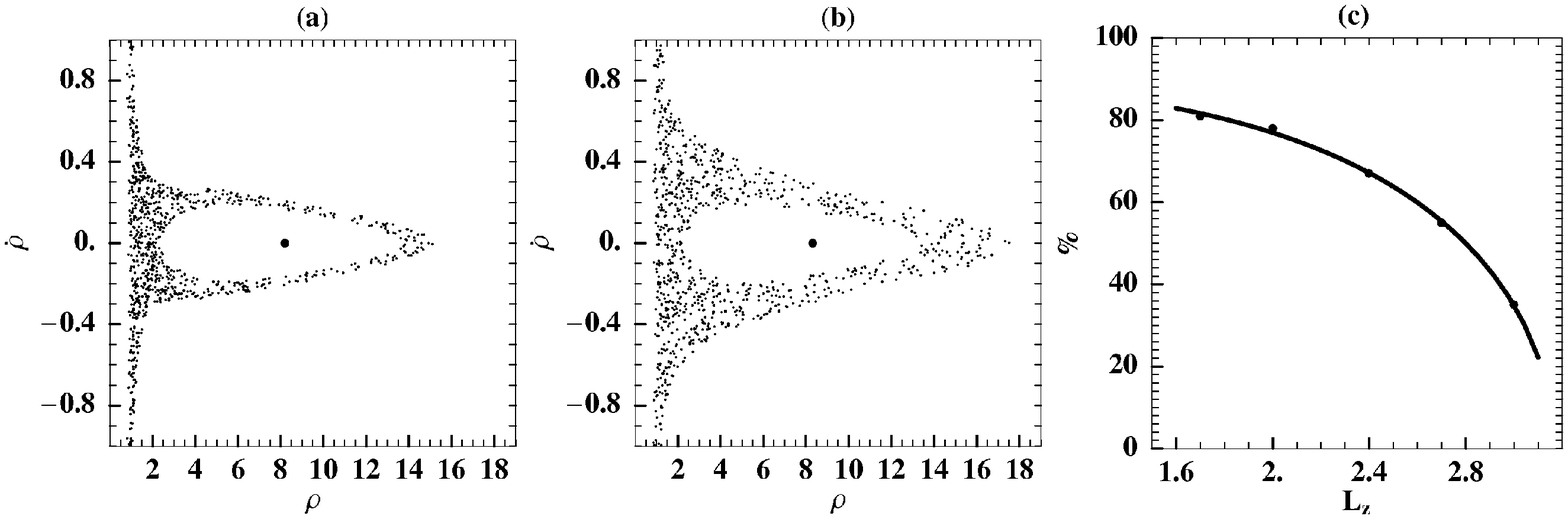,width=7.in} 
\end{center}
 \caption{Chaos on the Poincar\'{e} surfaces of section for (a) $L_z=2.7$ and
 (b) $L_z=1.7$, surrounding a large region containing ordered orbits (the
 closed blank area). (c) The proportion of the available phase space covered by
 chaotic orbits as a function of $L_z$ for $E=0.95,~q=0.95$.}
\label{FigChLz}
\end{figure}

While $L_z$ decreases (with constant $E=0.95$ and $q=0.95$) the proportion of
phase space surface occupied by the chaotic orbits on the surface of section
$z=0$ increases. For $L_z=3$ most of the chaotic orbits are inside the inner
closed CZV, but this region contains also an important island of stability (see
Fig.~\ref{FigIntReg}). There are also very small regions of chaos around the 
unstable periodic orbits of the outer permissible region. The proportion of the
area on the surface of section z=0 that is occupied by chaotic orbits is about
$20\%$. In order to estimate this fraction, we measured the area occupied by the
chaotic orbits on the surface of section and we divided it by the total area
covered by the allowed orbits. At lower values of $L_z$, the inner island of
stability (around $\mathbf{u}_0^\prime$) shrinks in size, and chaos  starts
occupying the envelope of the ordered region in the outer part of the now joined
permissible region (Fig.~\ref{FigLz2_98}b). The proportion of the  total area on
the surface of section $z=0$ occupied by chaotic orbits as a function of $L_z$
is given in Fig.~\ref{FigChLz}c. We see that the proportion of chaotic orbits
increases as $L_z$ decreases (Figs.~\ref{FigChLz}a,b), and for $L_z\leq 1.6$ it
is already larger than $80\%$.

\begin{figure}[ht]
\begin{center}
\psfig{file=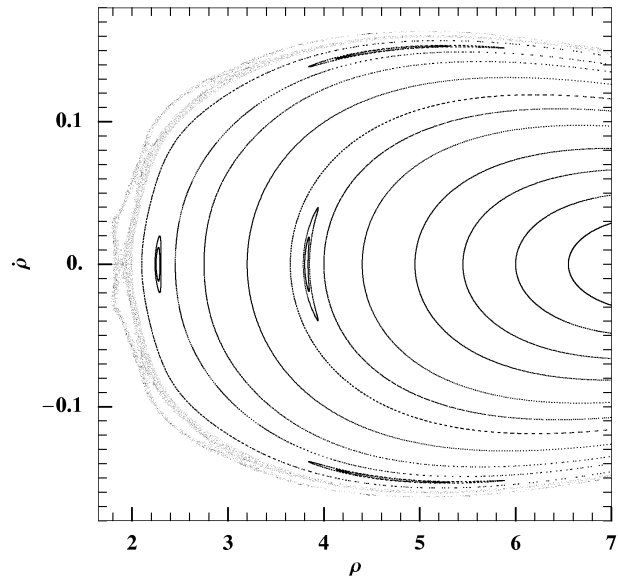,width=3.2in} 
\end{center}
 \caption{ A sticky chaotic domain (dark ribbon with holes) surrounding the
 region of ordered orbits for $L_z=0.1,~,E=0.95,~q=0.95$. In the ordered region
 there are islands of stability of the resonances $1/3$ and $1/2$. The sticky
 zone contains also several higher order islands of stability.}
\label{FigLz01}
\end{figure}

For even smaller values of $L_z$ ($L_z \lesssim 1.6$) all the chaotic orbits
belonging to the chaotic sea surrounding the main island of stability plunge
through the horizon, while the chaotic orbits which appear between the islands
inside the main island remain non-plunging. On the other hand orbits just
outside the boundary of the main island stay there for a long time before they
plunge in. These orbits that stay close to the boundary of the main island
exhibit stickiness (for a review about stickiness see \cite{Contop02}). In
Fig.~\ref{FigLz01} we see this sticky domain as a dark ribbon with small empty
holes. The empty holes contain higher order islands of stability. Orbits
starting on the left of the dark ribbon of Fig.~\ref{FigLz01} plunge in through
the horizon very fast and do not produce the densely populated chaotic regions
as in Figs.~\ref{FigChLz}a,b.

\begin{figure}[ht]
\begin{center}
\psfig{file=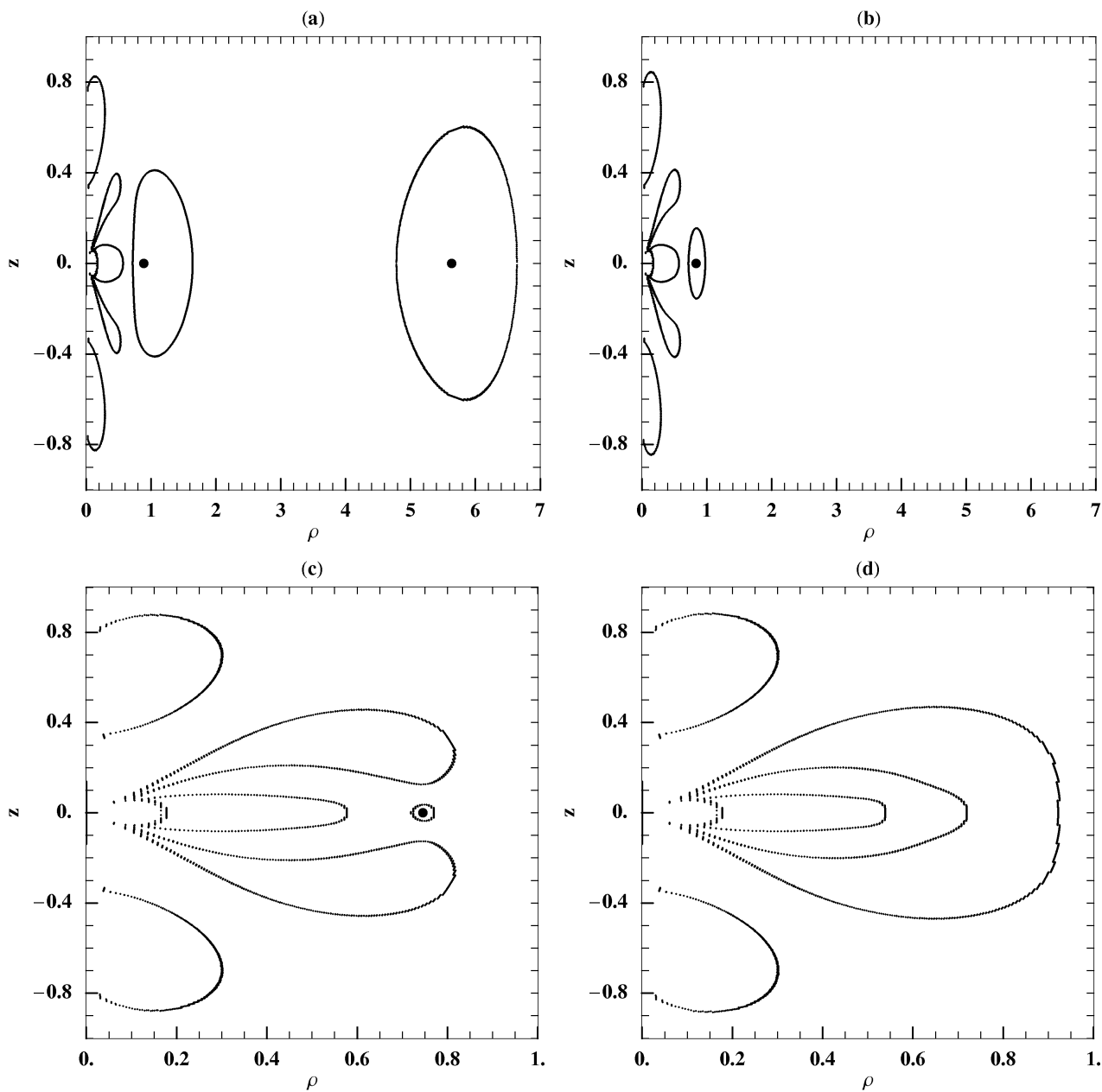,width=7.in} 
\end{center}
 \caption{ The CZVs for (a) $E=0.935$, (b) $E=0.7$, (c) $E=0.2$ and (d) $E=0.1$
 while $L_z=3,~q=0.95$.}
\label{FigRE}
\end{figure}

Next we consider the orbits for low values of $E$. In order to study them we
first examine how the shape of the CZVs changes if we keep $L_z$ and $q$
constant ($L_z=3$, $q=0.95$), but decrease the energy. What we get is the series
of curves of Fig.~\ref{FigRE}. As $E$ decreases below $E=0.95$ the two
permissible regions of non-plunging orbits get further and further apart and
they shrink. For $E\approx 0.93210$ the outer region disappears and for even
smaller $E$, there is only one permissible region with non-plunging orbits, the
one that was previously called the inner region. This region shrinks further as
$E$ decreases (compare Fig.~\ref{FigRE}a corresponding to $E=0.935$ and
Fig.~\ref{FigRE}b corresponding to $E=0.7$)  and for $E=0.2$ it is restricted
between the three central CZVs of plunging orbits (Fig.~\ref{FigRE}c). For even
smaller $E$ the two CZVs of plunging orbits, above and below the equatorial
plane join. When this happens most orbits plunge fast through the horizon
(Fig.~\ref{FigRE}d).

\begin{figure}[ht]
\begin{center}
\psfig{file=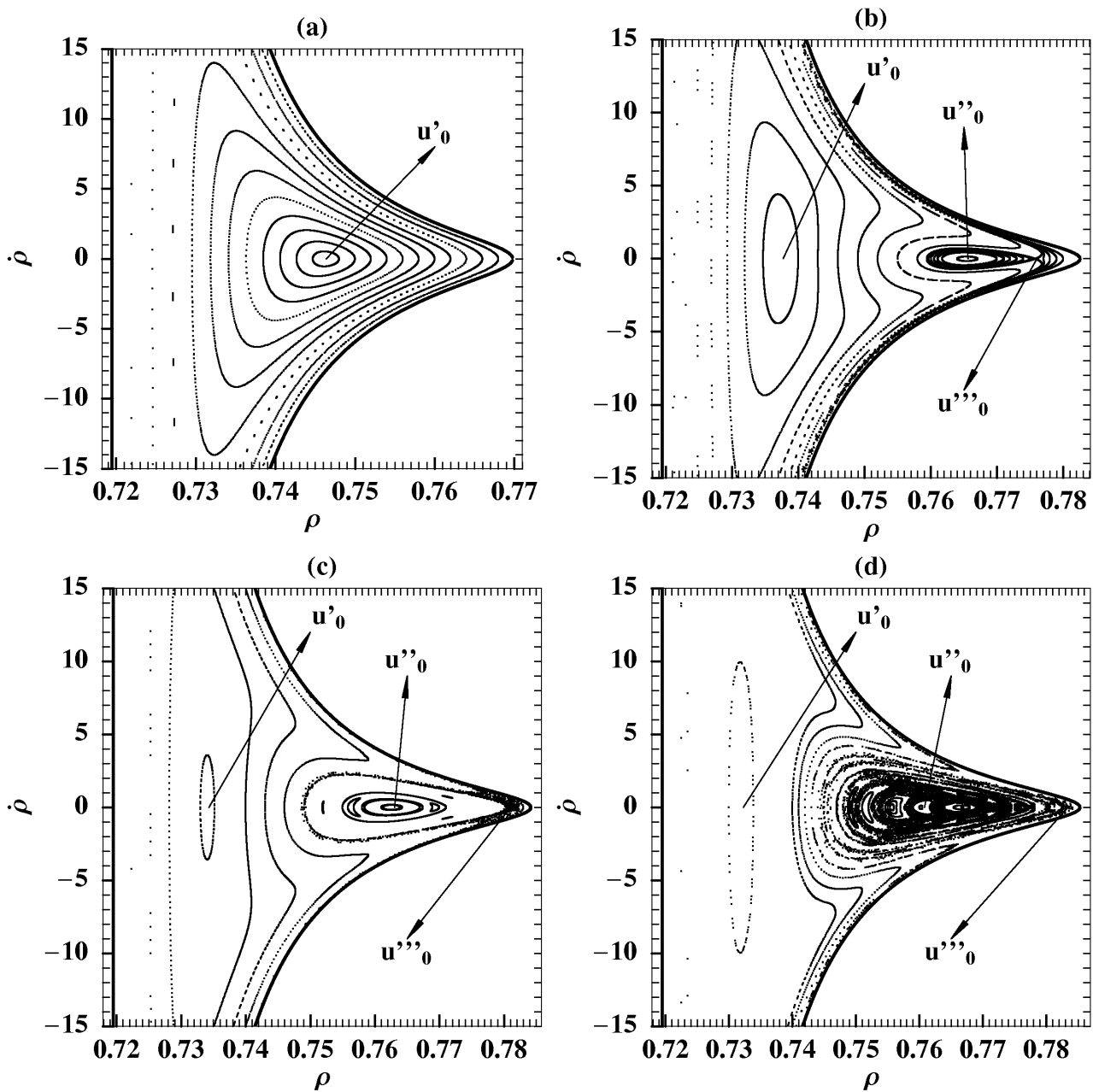,width=7.in} 
\end{center}
 \caption{Invariant curves and chaos on the surface of section
 $(\rho,\dot{\rho},z=0)$ for (a) $E=0.2$, (b) $E=0.192$, (c) $E=0.191$ and
  (d) $E=0.1903$ (with $L_z=3,q=0.95)$.}
\label{FigSSE}
\end{figure}

The details of the transition from Fig.~\ref{FigRE}c to \ref{FigRE}d are of
particular interest. The non plunging region of Fig.~\ref{FigRE}c for $E=0.2$
consists almost exclusively of ordered orbits, which are shown in
Fig.~\ref{FigSSE}a. The orbits are deployed around a central periodic orbit
$\mathbf{u}_0^\prime$ at $\rho \approx 0.745$. As $E$ decreases the periodic
orbit moves closer to the horizon, and for $E=0.192$ it is at
$\rho \approx 0.736$ (Fig.~\ref{FigSSE}b). In Fig.~\ref{FigSSE}b we see also
another stable periodic orbit (point $\mathbf{u}^{\prime\prime}_0$) at
$\rho\approx 0.764$ and an unstable periodic orbit at $\rho \approx 0.775$
(point $\mathbf{u}^{\prime\prime\prime}_0$). The latter pair of orbits was first
formed at $E \simeq 0.193$ at a ``tangent bifurcation'' (these orbits bifurcate
only from each other). Most orbits in Fig.~\ref{FigSSE}b are ordered and form
closed invariant curves. In particular the invariant curves close to the
boundary surround both islands around $\mathbf{u}^{\prime}_0$ and
$\mathbf{u}^{\prime\prime}_0$. There is only a very small chaotic region close
to the unstable point $\mathbf{u}^{\prime\prime\prime}_0$ and its asymptotic
curves.

As $E$ increases further  the chaotic region around the unstable point
$\mathbf{u}^{\prime\prime\prime}_0$ swells and diffuses towards the boundary of
the islands (Fig.~\ref{FigSSE}c). At particular values of $E$ we have
bifurcations of higher order periodic orbits from the central point
$\mathbf{u}^{\prime\prime}_0$. E.g. in Fig.~\ref{FigSSE}c (corresponding to
$E=0.191$) we see a double island with rotation number $\nu_\theta=1/2$ and a
triple island with rotation number $\nu_\theta=1/3$.

For $E=0.1903$ (Fig.~\ref{FigSSE}d) chaos has increased considerably and we see
further bifurcating islands around $\mathbf{u}^{\prime\prime}_0$ (now at
$\rho \approx 0.760$). In particular we see 2 islands with rotation number
$\nu_\theta = 1$ ($\rho \approx 0.756$ and $\rho \approx 0.765$); these two
islands consist of two distinct orbits, in contrast to what happens when
$\nu_\theta=1/2$ where an orbit starting at one island passes alternatively
through both islands.

\begin{figure}[ht]
\begin{center}
\psfig{file=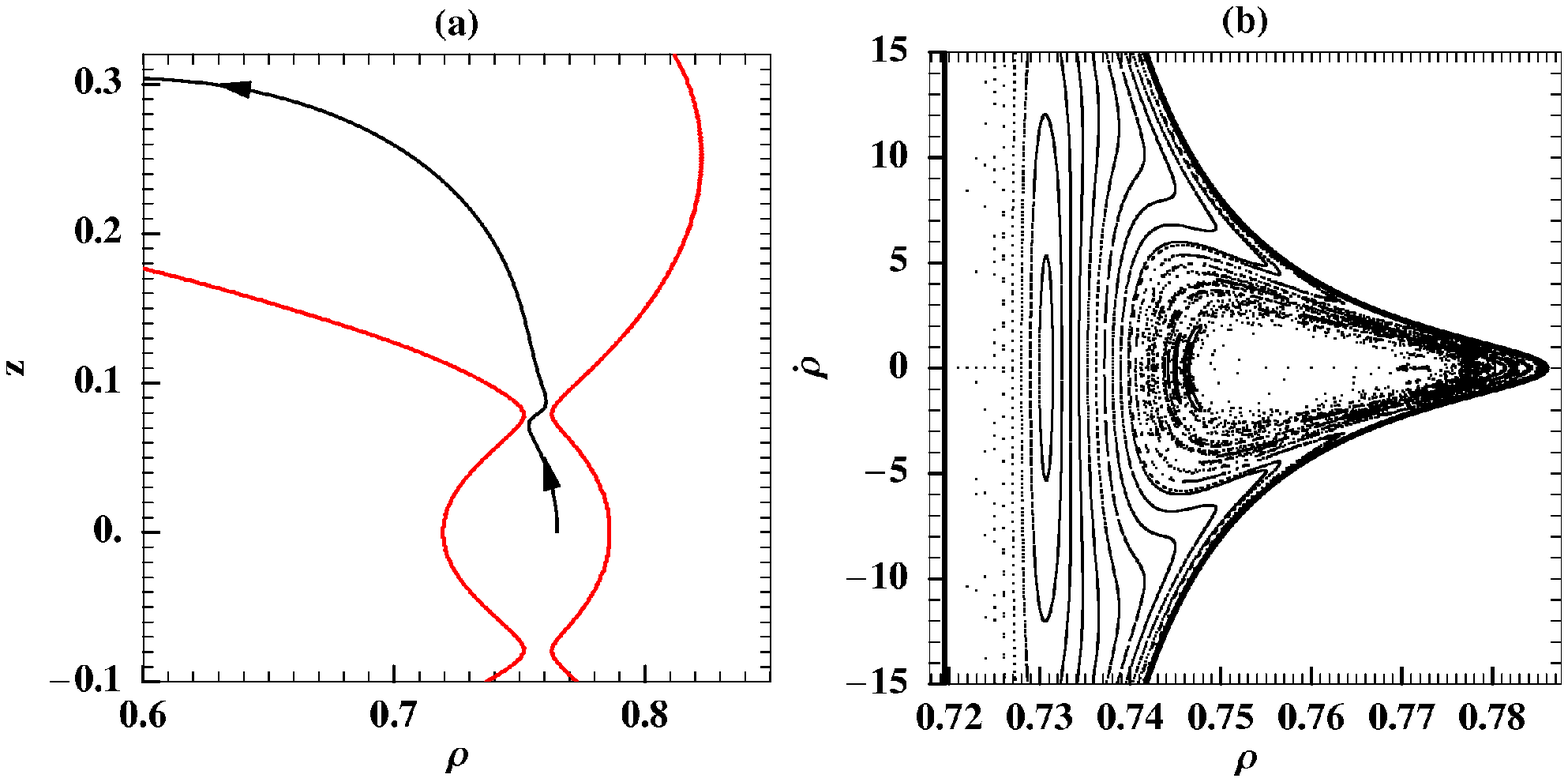,width=7.in} 
\end{center}
 \caption{(a) An orbit escaping upwards $(z>0)$ and plunging through the horizon
  for $E=0.1899$ $(L_z=3,q=0.95)$. (b) The corresponding surface of section.
  There is a large island surrounding $\mathbf{u}^{\prime}_0$ (near
  $\rho=0.7308$, $\dot{\rho}=0$), some islands of higher order on the right
  part of the figure, and an almost blank region of the fast escaping orbits,
  surrounded by chaotic orbits that escape after a larger time.}
\label{FigEsc}
\end{figure}

When $E < 0.1901$ the lobes of the curves of zero velocity above and below the
region of the non-plunging region of Fig.~\ref{FigRE}c join the central
non-plunging region. Then most of the chaotic orbits escape upwards or downwards
along the z-axis, and finally plunge through the horizon (see Fig.~\ref{FigEsc}a 
which corresponds to $E=0.1899$). However, an island of stability still remains 
around $\rho = 0.73$ (Fig.~\ref{FigEsc}b). As $E$ decreases further the stable
periodic orbit $\mathbf{u}^\prime_0$ moves further inwards, very close to the
inner boundary. For $E <0.1858$ $\mathbf{u}^\prime_0$ becomes unstable and for a 
little smaller $E$ the island around it disappears. On the other hand the island 
generated around $\mathbf{u}^{\prime\prime}_0$ persists for $E=0.18$ but for 
$E=0.17$ it has disappeared.

It is remarkable that the inner permissible region of the non-plunging orbits
assumes its smallest size for $E\approx 0.265$, below which it starts expanding
again before it gets swallowed by the lobes of the plunging regions.

\begin{figure}[ht]
\begin{center}
\psfig{file=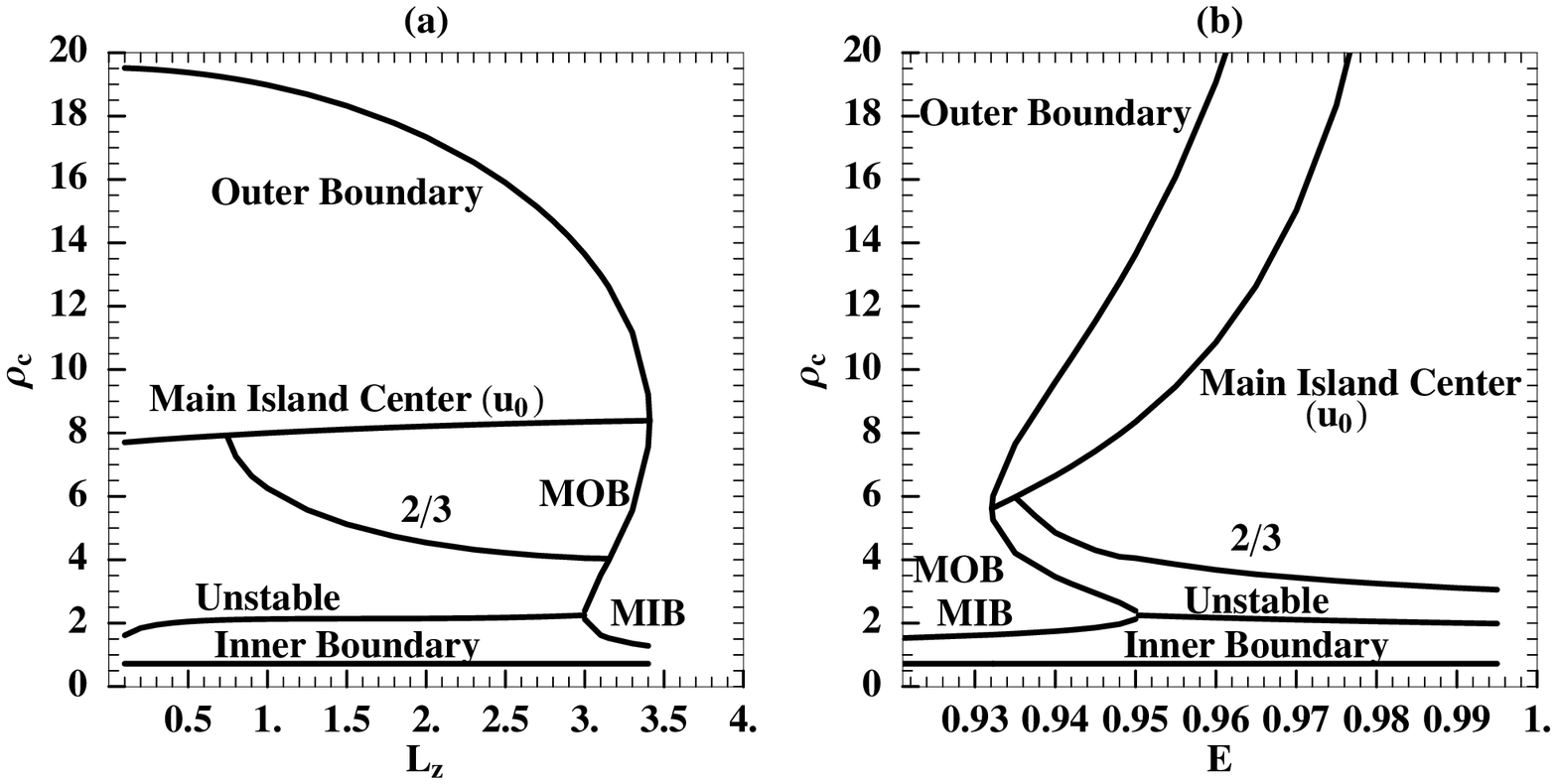,width=7.in} 
\end{center}
\caption{The characteristics of the main families of periodic orbits (for
$q=0.95$, $\chi=0.9$) in the MN metric along the line $\rho$ on the surface of
section at $z=0$ and the corresponding boundaries of the permissible regions
(a) as functions of $L_z$ for $E=0.95$, and (b) as functions of $E$ for
$L_z=3$.} 
\label{FigPosM}
\end{figure}

\section{Characteristics} \label{sec:Charact}

The position of a periodic orbit as a function of a parameter of the system is
called a characteristic. In Fig.~\ref{FigPosM} we give the characteristics of
the main families of periodic orbits  in the MN metric (for $q=0.95$,
$\chi=0.9$), namely the central periodic orbit of the outer region (the ``main
island center'' $\mathbf{u}_0$), the periodic orbit $2/3$ (the point lying on
the line $\dot{\rho}=0$ of the surface of section at $z=0$), and the unstable 
saddle orbit, as functions of $L_z$ for $E=0.95$ (Fig.~\ref{FigPosM}a) and as 
functions of $E$ for $L_z=3$ (Fig.~\ref{FigPosM}b). In the same figures we have
drawn also the boundaries of the permissible regions for $z=0$, i.e.~the 
intersections of the CZVs with the axis $z=0$. These are the outer and the inner 
boundaries of the non-plunging orbits region, when the non-plunging regions are
joined. When there are two distinct regions containing non-plunging orbits we
mark the middle outer boundary (MOB), that is the inner boundary of the outer 
region, and the middle inner boundary (MIB), that is outer boundary of the inner 
region.

In Fig.~\ref{FigPosM}a we see that the central orbit ($\mathbf{u}_0$) has an
almost constant position, while the orbit $2/3$ bifurcates from the central
family for $L_z\approx 0.75$ and moves inwards (towards $\rho=0$) as $L_z$
increases. This orbit reaches the MOB for $L_z\approx 3.15$ while for larger
$L_z$ $\mathbf{u}_0$ does not exist at all. The saddle point is generated when
the two non-plunging CZVs merge for $L_z = 2.99762$ and then it moves inwards as
$L_z$ decreases. On the other hand the outer boundary of the outer CZV moves to
large distances when $L_z$ decreases, while the corresponding inner boundary
remains at a constant distance from the horizon. The MOB and the MIB show up for
$L_z \geq 2.99762$. For larger $L_z$ the MIB and the MOB get further apart,
i.e.~the MIB moves inwards while the MOB moves outwards. For
$L_z \approx 3.40935$ the outer boundary and the MOB join, and then the outer
region completely disappears. On the other hand the MIB moves inwards as $L_z$
increases and appears to level off for $L_z>3.4$.

In Fig.~\ref{FigPosM}b we give the characteristics and boundaries as functions
of $E$. As $E$ increases, the outer boundary increases enormously, approaching
infinity as $E\rightarrow 1$, while the inner boundary remains at an almost 
constant position. The central periodic orbit ($\mathbf{u}_0$) moves also
outwards for increasing $E$, while the orbit $2/3$ moves inwards. The orbit
$2/3$ bifurcates from the central orbit for $E \approx 0.935$. For large $E$ we
have a unique large CZV containing the non plunging orbits, but for
$E \leq 0.95038$ the permissible region splits into two. The inner region is
between the MIB and the inner boundary. This region shrinks slightly as $E$
decreases. The outer region is between the MOB and the outer boundary. This
region shrinks quickly as $E$ decreases and finally at $E \approx 0.93210$ it
disappears. For $E \leq 0.93210$ there is no outer region and thus no central
periodic orbit $\mathbf{u}_0$. As long as the CZV is unique ($E \geq 0.95038$)
there is also an unstable saddle point which moves slightly inwards as $E$
increases.

\begin{figure}[ht]
\begin{center}
\psfig{file=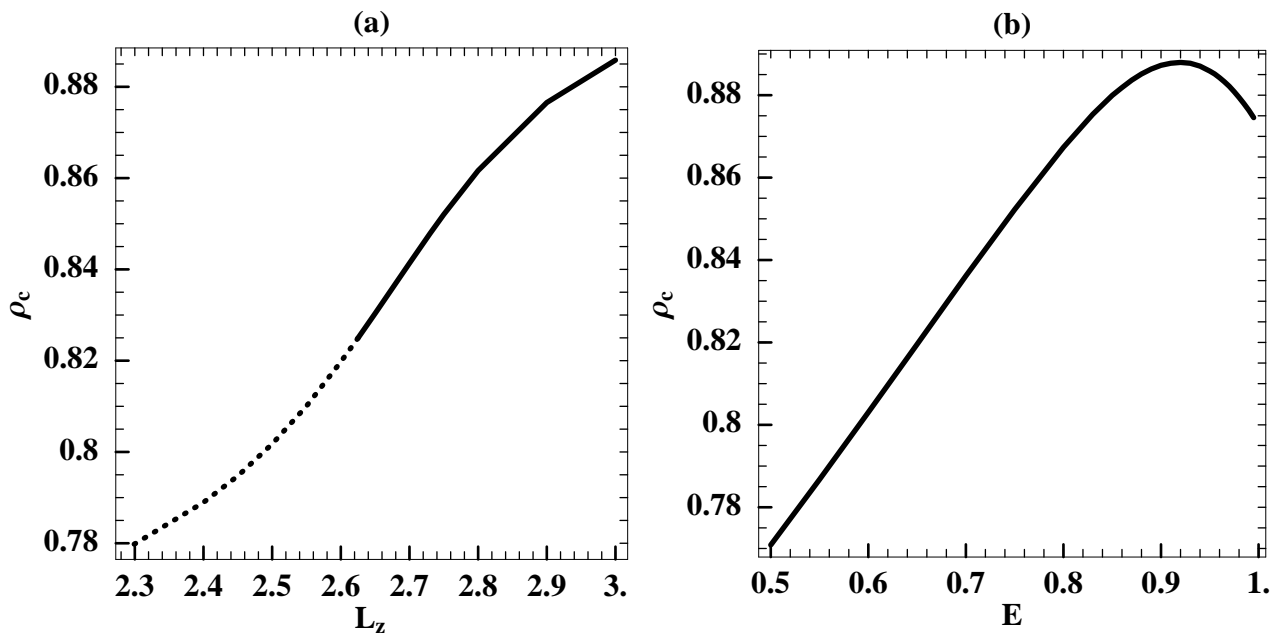,width=7.in} 
\end{center}
 \caption{The characteristic of the periodic orbit $\mathbf{u}^\prime_0$ lying
 in the center of the main island of the inner region (for $q=0.95$, $\chi=0.9$)
 (a) as function of $L_z$ for $E=0.95$ and (b) as function of $E$ for
 $L_z=3$. The dotted lower part of the curve of Fig.~\ref{FigInSt}a indicates
 that the periodic orbit has become unstable.}
\label{FigInSt}
\end{figure}

\begin{figure}[ht]
\begin{center}
\psfig{file=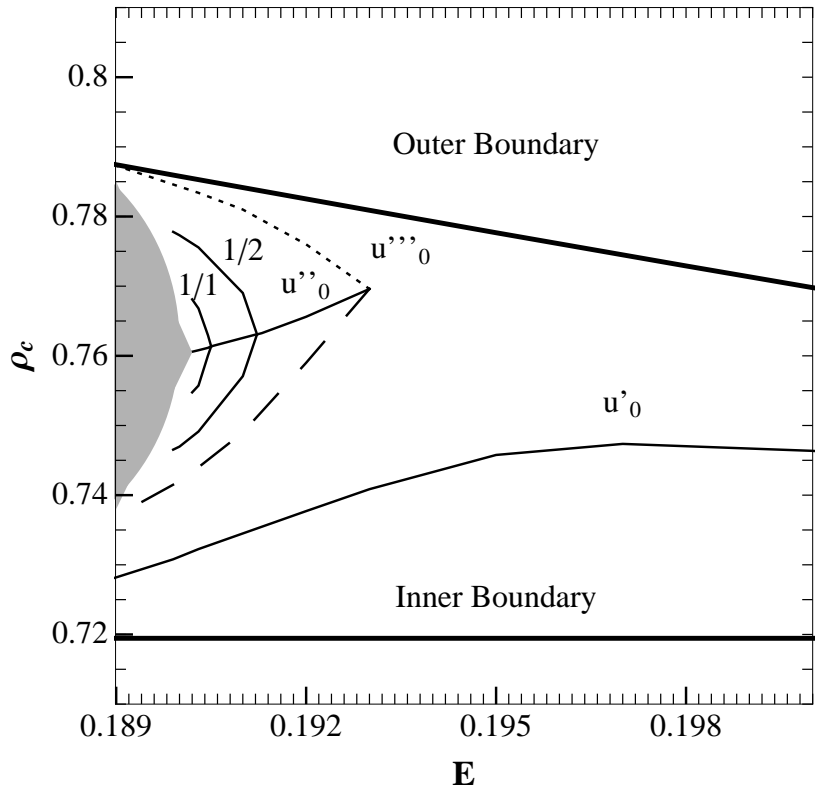,width=4.5in} 
\end{center}
 \caption{Characteristics of the periodic orbits and boundaries of the inner
 region for $0.189<E<0.2$. Stable orbits are denoted with continuous lines, and
 unstable orbits with dotted lines. The unstable intervals of
 $\mathbf{u}^{\prime\prime}_0$ on the left of the bifurcating families $1/2$ and
 $1/1$ are very small and they are not marked. The gray area for $E \leq 0.1901$
 represents chaotic orbits plunging through the horizon. The asymptotic curves
 of the unstable orbit $\mathbf{u}^{\prime\prime\prime}_0$ intersect at the
 first homoclinic points along the dashed curve. Close to this curve and to the
 curve $\mathbf{u}^{\prime\prime\prime}_0$ the orbits are chaotic.}
\label{FigERI}
\end{figure}

The characteristics of the stable periodic orbit $\mathbf{u}_0^\prime$ of the
inner region as functions of $L_z$ and $E$ are shown in Fig.~\ref{FigInSt}. The
periodic orbit $\mathbf{u}_0^\prime$ is very close to the inner boundary shown
in Fig.~\ref{FigPosM}. For $E=0.95$ and $L_z > 2.6125$, $\mathbf{u}_0^\prime$ is 
stable, but for $L_z < 2.6125$ it becomes unstable. On the other hand as we
decrease $E$ for constant $L_z=3$, the periodic orbit $\mathbf{u}_0^\prime$
moves initially (as long as $E \gtrsim 0.92$) outwards (see Fig.~\ref{FigInSt}b), 
but for $E \lesssim 0.92$, $\mathbf{u}_0^\prime$ moves inwards, until it reaches
a minimum value for $E\approx 0.265$. For the interval
$0.265 \gtrsim E \gtrsim 0.196$ the point $\mathbf{u}_0^\prime$ moves slightly 
outwards for decreasing $E$, and then for $E \lesssim 0.196$, 
$\mathbf{u}_0^\prime$ moves inwards again (Fig.~\ref{FigERI}). The size of the 
region of permissible motion decreases as $E$ decreases (Fig.~\ref{FigRE}a,b),
but for $E \lesssim 0.265$ this region starts expanding again.

The details of the characteristics of the periodic orbits for $E\leq 0.2$ are
shown in Fig.~\ref{FigERI}. We see the characteristic of the main periodic orbit
$\mathbf{u}^\prime_0$ of the inner island and the boundary of the permissible
region along the line $z=0$. Furthermore we have drawn the characteristics of
the periodic orbits $\mathbf{u}^{\prime\prime}_0$ and 
$\mathbf{u}^{\prime\prime\prime}_0$ of Fig.~\ref{FigSSE}, and of the main 
bifurcating families from $\mathbf{u}^{\prime\prime}_0$ along the axis
$\dot{\rho}=0$. The rotation number along every family bifurcating from 
$\mathbf{u}^{\prime\prime}_0$ is constant and it is marked in Fig.~\ref{FigERI}. 
The rotation number of the orbit $\mathbf{u}^{\prime\prime}_0$ (the limit of the 
rotation number of the invariant curves around $\mathbf{u}^{\prime\prime}_0$ as 
they shrink to $\mathbf{u}^{\prime\prime}_0$) increases as E decreases. When the 
rotation number $\nu_\theta$ of $\mathbf{u}^{\prime\prime}_0$  becomes 
$\nu_\theta=1/2$ the orbit $\mathbf{u}^{\prime\prime}_0$ becomes unstable and a 
stable family $1/2$ bifurcates. For a small interval of values $\Delta E$ of 
decreasing $E$, $\mathbf{u}^{\prime\prime}_0$ remains unstable, and for even 
smaller $E$ it becomes stable again. When $\nu_\theta=1$ the orbit 
$\mathbf{u}^{\prime\prime}_0$ generates by bifurcation a stable family $1/1$ and 
becomes again unstable. For smaller $E$ the family $\mathbf{u}^{\prime\prime}_0$ 
has an infinity of transitions to stability and instability. The phenomenon of 
infinite transitions to instability and stability that leads to the termination
of a family of periodic orbits at the escape energy was described in detail in a
different dynamical system \cite{Contop80,Contop87}. In the present case when
$E$ tends to $0.1901$ the period of the orbit $\mathbf{u}^{\prime\prime}_0$
should tend to infinity and for $E \leq 0.1901$, when the CZVs have opened, the
periodic orbit $\mathbf{u}^{\prime\prime}_0$ should not exist anymore. The
orbits close to $\mathbf{u}^{\prime\prime}_0$ escape and plunge through the
horizon as shown in Fig.~\ref{FigEsc}a. The region of escaping orbits is shown
in Fig.~\ref{FigERI} with gray color. We notice that this region increases
abruptly as $E$ decreases below the escape energy $E=0.1901$. 

In Fig.~\ref{FigERI} we have marked the characteristic of the unstable family
$\mathbf{u}^{\prime\prime\prime}_0$ (dotted line) and the locus of the first
homoclinic intersection of the asymptotic curves of the orbit
$\mathbf{u}^{\prime\prime\prime}_0$ (dashed line). The characteristic of
$\mathbf{u}^{\prime\prime\prime}_0$ approaches the outer boundary as $E$
decreases. Close to these lines the orbits are chaotic. However chaos is quite
limited for $E$ only slightly smaller than its value at the tangent bifurcation
of the orbits $\mathbf{u}^{\prime\prime}_0$ and
$\mathbf{u}^{\prime\prime\prime}_0$ (e.g.~for $E=0.192$, Fig.~\ref{FigSSE}b).
The asymptotic curves from the unstable orbit
$\mathbf{u}^{\prime\prime\prime}_0$ separate the invariant curves that close
around $\mathbf{u}^{\prime\prime}_0$ from those that close around
$\mathbf{u}^{\prime}_0$ (Fig.~\ref{FigSSE}b). As $E$ decreases chaos
becomes more important (Figs.~\ref{FigSSE}c,d), and when $E$ is smaller than
the escaping energy (the energy at which the closed CZV of (Fig.~\ref{FigSSE}c)
join the upper and lower CZVs of the plunging orbits) most chaotic orbits escape.

The family $\mathbf{u}^{\prime}_0$ remains stable as $E$ decreases to the left
of the border of Fig.~\ref{FigERI}, until $E \approx 0.186$. For a little
smaller $E$ this family becomes unstable and for even smaller $E$ the
surrounding invariant curves disappear. On the other hand the island formed by
the invariant curves around $\mathbf{u}^{\prime\prime}_0$ continues to exist,
even when this orbit and the nearby orbits escape, but for $E \lesssim 0.17$
almost all orbits escape.

\section{Nongeodesic orbits} \label{sec:NGO}

In real EMRI systems the energy and the angular momentum are not conserved;
they change adiabatically due to gravitational radiation. Thus the real orbits
are nongeodesic, although their deviation from geodesics is very small. An
estimate of the loss of energy and angular momentum per unit time for a test
body in Kerr metric, $\frac{dE}{dt}$ and $\frac{dL_z}{dt}$, has been provided by
\cite{Gair06} (see Eqs.~(44,45)). We use these formulae, appropriately modified,
to compute approximately the corresponding losses in our MN system
\cite{Gair08,Lukes10}.

For a relatively small time interval we may consider $\frac{dE}{dt}$ and
$\frac{dL_z}{dt}$  almost constant; thus
\begin{equation}
 E=E_0+\frac{dE_0}{dt}~t,~~~~~L_z=L_{z0}+\frac{d L_{z0}}{dt}~t
\end{equation}
where $E_0$, $L_{z0}$ are the corresponding initial values and $\frac{dE_0}{dt}$,
$\frac{d L_{z0}}{dt}$ are negative, and represent the constant rates of loss.
The rates of loss depend on the mass ratio of the EMRI system; in fact as this
ratio tends to $0$ the motion tends to become more and more geodesic. 

We have applied these approximate formulae in our numerical calculation of the
orbits for several orbital periods. During this time the structure of the phase
space changes gradually, e.g.~the location and the size of the resonant islands
change. If an orbit starts away from the main resonant islands it moves
adiabatically along different invariant curves and, as a consequence, its
rotation number gradually changes in an apparently strictly monotonic way. But
if the orbit passes through an island of stability its rotation number remains
constant and equal to a fixed rational number, like $2/3$.

In the corresponding integrable case (the Kerr metric), there are no islands of
stability, thus the variation of the rotation number $\nu_\theta$ is always
strictly monotonic in time. However, in a non-integrable non-Kerr metric (like
the MN metric) there are several islands of stability and an orbit passing
through them assumes a constant value of $\nu_\theta$ for some interval of time.
In this interval the rotation curve exhibits a corresponding plateau. The
existence of such plateaus can be used as an observational criterion to
distinguish a Kerr from a non-Kerr background, since the rotation number can be
inferred from the spectrum of the gravitational waves.

\begin{figure}[ht]
\begin{center}
\psfig{file=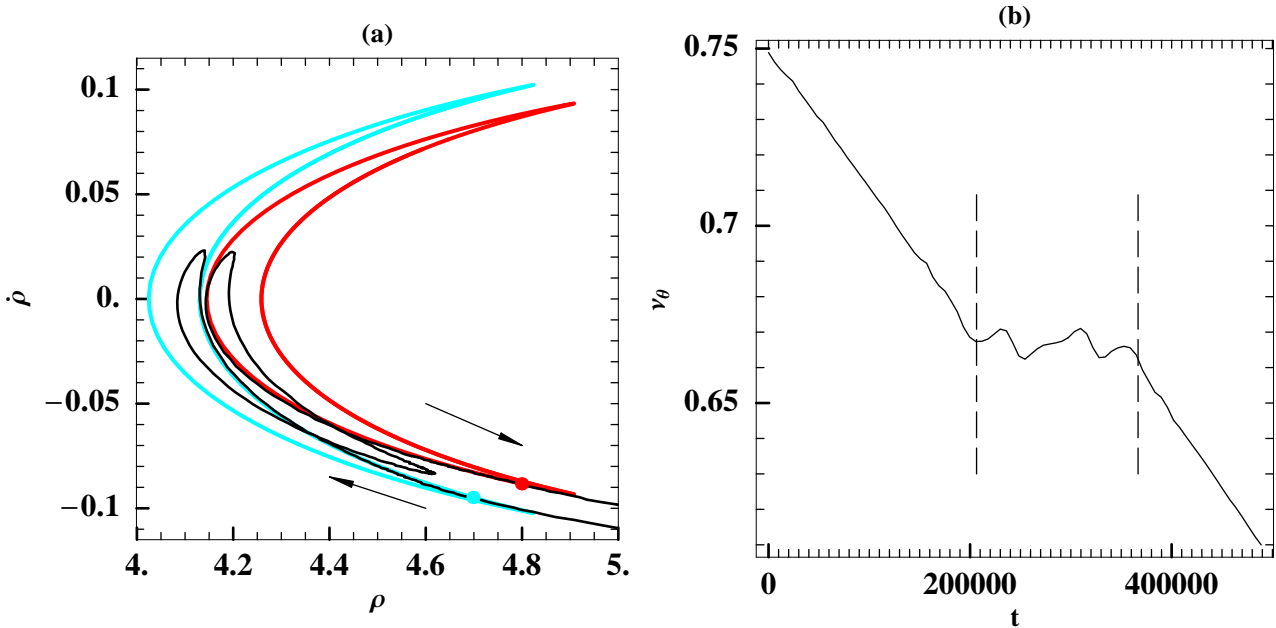,width=7.in} 
\end{center}
 \caption{(a) The black line joins the third intersections of a non geodesic
 orbit ($z=0$) and passes through an island of stability of the $2/3$ resonance.
 This black line performs a couple of windings on a surface of section while the
 non geodesic loses energy and angular momentum due to gravitational radiation. 
 The blue point indicates the point of entrance of the nongeodesic orbit to the
 resonance, while the red indicates the exit. The blue and the red curves
 represent the geodesic orbits if the blue point and the red point are used as
 an initial conditions. (b) The evolution of the ratio of the corresponding
 fundamental frequencies as a function of time for the orbit in (a). The dashed
 lines show the time interval during which the non geodesic orbit is trapped.
 The present evolution of the non geodesic orbit corresponds to a binary with
 ratio of masses equal to $8 \times 10^{-5}$.}
\label{FigNGOr}
\end{figure}

In Fig.~\ref{FigNGOr}a we show a non geodesic orbit for an interval of time
during which the orbit crosses the resonance $2/3$. Initially the non geodesic
orbit is outside but close to the island of stability $2/3$ and its initial 
condition lies in the area between the $2/3$ resonance and the central orbit 
$\mathbf{u}_0$. The angles formed between the successive intersections of the
non geodesic orbit on the surface of section at $z=0$ as they are seen from the 
central periodic orbit $\mathbf{u}_0$ are slightly larger than $4 \pi/3$. Thus 
third intersection lies approximately at a particular invariant curve with
$\nu_\theta>2/3$. In Fig.~\ref{FigNGOr}a we have joined every third intersection
by a continuous line. Thus the non geodesic curve crosses successive invariant
curves and revolves around $\mathbf{u}_0$. When these intersections enter the 
$2/3$ resonance, e.g.~the leftmost island of the 3 islands 
(Fig.~\ref{FigPSLz3E095}), they tend to follow the circulation of the geodesic 
orbits, corresponding to the instantaneous values of $E$ and $L_z$, that form
the island itself. As time progresses and $E,~L_z$ change, these resonance
islands move gradually closer to $\mathbf{u}_0$, e.g.~the leftmost island moves
to the right. The non geodesic orbit follows the drift of the island while it
moves around the island for some time. If the time it remains inside the island  
is sufficient, each third intersection of the non geodesic orbit may form
a number of loops. In Fig.~\ref{FigNGOr}a they form 2 loops inside the leftmost 
island. Later on these intersections exit the island and then form a curve which 
approximates invariant curves around the central orbit $\mathbf{u}_0$ with
rotation number $\nu_\theta<2/3$. Because of the previously described relative  
``motion'' of the non geodesic orbit and the geodesic background, a non geodesic 
orbit whose initial condition lies between the $2/3$ resonance and the orbit 
$\mathbf{u}_0$ will eventually cross the $2/3$ resonance and will finally find 
itself on the other side of the resonance island.

The corresponding fundamental frequencies of the spectrum of the gravitational
radiation change in time and their ratio follows a curve like the one in
Fig.~\ref{FigNGOr}b.  When the non geodesic orbit is outside the resonance
($\nu_\theta>2/3$) the value of $\nu_\theta$ appears to decrease strictly
monotonically. When the non geodesic orbit is entrapped inside the resonance
its rotation number remains theoretically constant. After the non geodesic orbit
exits the resonant region the rotation number appears to decrease again strictly
monotonically. (For a more detailed discussion of the relation between the 
fundamental frequencies of the spectrum and $\nu_\theta$ see \cite{Lukes10}.)

In Fig.~\ref{FigPSLz3E095}b we see small oscillations of $\nu_\theta$ during the
trapping period. This is due to the finite time for which we evaluate the
frequencies during the evolution of the non geodesic loops. We cannot evaluate
the frequencies for larger time intervals, because the frequencies change
continuously, although this change is very slow. This imposed finiteness in the
frequency analysis has a side effect: the loops are not averaged and therefore
an extra illusive frequency -the one connected with the circulation of the
island-appears.

\section{Conclusions} \label{sec:Conc}

In this paper we have studied thoroughly the details of the various types of
orbits in a MN spacetime. The non-integrability of the corresponding geodesic
equations is quite clear since all the general characteristics exhibited by
systems that deviate from an integrable one are present in this case as well.
Namely, more than one chains of Birkhoff islands are present in a surface of
section, while chaotic regions are also present. The most important chaotic
region is mainly present in the inner closed CZV, where the deviation from the
Kerr metric is more pronounced. Also when the two regions of allowed orbits are
joined, the chaotic behavior of the inner region is partly transferred in the
outer region. We have also examined the behavior of the regions containing non
plunging orbits when the parameters of the system (mainly $E$, $L_z$) change,
and take extreme values.

Finally we studied the case of a non geodesic orbit as it moves adiabatically
in and out of a resonance island. The plateau in the evolution of the ratio of
the corresponding gravitational wave signal frequencies, that was analyzed in
\cite{Lukes10}, is present and can be clearly used as a tool to distinguish
a Kerr from a non-Kerr metric.

\end{document}